\documentstyle[aps,preprint]{revtex}
\begin{document}
\draft
\author{Yi-shi Duan and Peng-ming Zhang\footnote{Anthor to whom
correspondence should be addressed. Email: zhpm@lzu.edu.cn}}
\title{Decomposing the $SU(N)$ connection and the Wu-Yang potential}
\date{\today}
\address{Institute of Theoretical Physics, Lanzhou University, Lanzhou,
730000, People' Republic of China}
\maketitle

\begin{abstract}
Based on the decomposition of $SU(2)$ gauge field, we derive a
generalization of the decomposition theory for the $SU(N)$ gauge field. We
thus obtain the invariant electro-magnetic tensors of $SU(N)$ groups and the
extended Wu-Yang potentials. The sourceless solutions are also discussed.
\end{abstract}

\pacs{PACS numbers: 11.15.Tk, 02.40.-k}

%\maketitle

\section{Introduction}

Recently Faddeev et al. proposed a decomposition of the four
dimensional $SU(N)$ Yang-Mills field $A_{\mu }^{a}$ \cite
{Faddeev,Shabanov,Cho}. In their paper with some ansatz the new
variables were given for studying the knot theory and the QCD. One
of our author (Duan) pointed out that the gauge potential should
be decomposed in terms of the gauge covariant \cite{DuanGe}, i.e.
$A_{\mu }=a_{\mu }+b_{\mu }$, which the $a_{\mu }$ satisfies the
gauge transformation $a_{\mu }^{\prime }=ga_{\mu }g^{-1}+\partial
_{\mu }gg^{-1}$ and the $b_{\mu }$ satisfies the adjoint
transformation $b_{\mu }^{\prime }=gb_{\mu }g^{-1}.$ The $a_{\mu
}$ part may show the geometry property of system and the $b_{\mu
}$ part may be looked upon as vector boson which would be massive.
Based on the decomposition theory of the $SU(2)$ gauge field
\cite{DuanGe}, the gauge potential of the $SU(N)$ gauge field is
decomposed in terms of local bases corresponding to the Cartan
subalgebra without any hypothesis. With this decomposition the
extended 't Hooft electromagnetic tensor is derived and the
Wu-Yang potential is given. At last, we discuss the sourceless
solution of the gauge field equation of group $SU(N).$

\section{SU(2) gauge field and the 't Hooft monopole}

In this section we introduce the decomposition theory of SU(2)
gauge field and corresponding Wu-Yang potential. In terms of the
$SU(2)$ gauge theory, the covariant derivation of a unit gauge
field $n^a(x)$ is
\[
D_\mu n^a(x)=\partial _\mu n^a(x)+\varepsilon ^{abc}B_\mu
^bn^c,\;\;a,b,c=1,2,3,
\]
where $B_\mu ^a$ is the gauge potential of the $SU(2)$ gauge theory. By
virtue of this definition we can give the expression of the gauge potential $%
B_\mu ^a$
\begin{equation}
B_\mu ^a=A_\mu n^a+\varepsilon ^{abc}\partial _\mu n^bn^c-\varepsilon
^{abc}D_\mu n^bn^c,  \label{dec1}
\end{equation}
where $A_\mu =B_\mu ^an^a$ is the Abelian projection of the $SU(2)$ gauge
potential. Thus the gauge potential is decomposed formally, one will find
that the decomposition is very useful in the 't Hooft monopole theory. We
calculate the corresponding field strength
\[
F_{\mu \nu }=\partial _\mu B_\nu -\partial _\nu B_\mu -[B_\mu
,B_\nu ],
\]
from the Eq. (\ref{dec1}) we give
\begin{equation}
F_{\mu \nu }^an^a=(\partial _\mu A_\nu -\partial _\nu A_\mu
)-\varepsilon ^{abc}n^a\partial _\mu n^b\partial _\nu
n^c+\varepsilon ^{abc}n^aD_\mu n^bD_\nu n^c.  \label{pro2}
\end{equation}
Since the left hand $F_{\mu \nu }^an^a$ and the last term of right
hand are the gauge invariant term, we find that the following
quantity is a gauge invariant
\[
f_{\mu \nu }=(\partial _\mu A_\nu -\partial _\nu A_\mu )-K_{\mu
\nu },
\]
where $K_{\mu \nu }=\varepsilon ^{abc}n^a\partial _\mu n^b\partial _\nu n^c$%
. One can find that $f_{\mu \nu }$ is just the gauge field tensor
defined by 't Hooft\cite{tHooft} which is fundamental for the
magnetic monopole
\begin{equation}
f_{\mu \nu }=F_{\mu \nu }^an^a-\varepsilon ^{abc}n^aD_\mu n^bD_\nu
n^c. \label{fie3}
\end{equation}

Let $\vec{e}_a(x)$ $(a=1,2)$ are the vierbein perpendicular to $\vec{n}(x),$
it is easy to verify that there exists a $U(1)$ potential
\begin{equation}
a_\mu =\varepsilon ^{ab}(\vec{e}_a\cdot \partial _\mu \vec{e}_b),
\label{dec3}
\end{equation}
which satisfies $K_{\mu \nu }=\partial _\mu a_\nu -\partial _\nu
a_\mu .$ One can find that this potential is just the Wu-Yang
potential\cite{Wu1,Wu2} of the magnetic monopole system.

\section{Local basis and the SU(N) connection}

Let $T_a$ ($a=1,2,...,r)$ be Lie algebraic bases of the $SU(N)$ group $G$,
and $H_i(i=1,2,...,L)$ the Cartan subalgebra, i.e.
\begin{equation}
\lbrack T_a,T_b]=if_{abc}T_c,\;\;[H_i,H_j]=0.  \label{com1}
\end{equation}
The local basis of Cartan subalgebra is defined as
\begin{equation}
n_i(x)=U(x)H_iU^{\dagger }(x),  \label{loc1}
\end{equation}
where $U(x)$ is a unitary matrix on manifold $M.$ The covariant derivative
of the local basis $n_i(x)$ is
\begin{equation}
D_\mu n_i=\partial _\mu n_i-ig[A_\mu ,n_i]  \label{Dn}
\end{equation}
where $A_\mu $ is a $su(N)$ Lie algebra vector
\[
A_\mu =A_\mu ^aT_a.
\]
The curvature is defined as
\begin{equation}
F_{\mu \nu }=\partial _\mu A_\nu -\partial _\nu A_\mu -ig[A_\mu ,A_\nu ].
\end{equation}
In terms of the relation
\begin{equation}
f_{abl}f_{acm}n_i^bn_j^c+n_i^ln_j^m=\delta ^{lm},
\end{equation}
one can find that a $su(N)$ Lie algebra vector $V$ can be decomposed with
the local basis
\begin{equation}
V=(V,n_i)n_i+[[V,n_i],n_i],\;\;\;\;\forall V\in su(N).
\end{equation}
Since it is a $su(N)$ Lie algebra vector, with the Eq. (\ref{Dn}) the $SU(N)$
connect $A_\mu $ can be decomposed as
\begin{equation}
A_\mu =(A_\mu ,n_i)n_i+\frac 1{ig}[\partial _\mu n_i,n_i]-\frac 1{ig}[D_\mu
n_i,n_i].  \label{con-de}
\end{equation}
Similiar to the 't Hooft electromagnetic tensor in the $SU(2)$ gauge field,
we can define the extended 't Hooft electromagnetic tensor
\begin{equation}
f_{\mu \nu }^i=(F_{\mu \nu },n_i)+\frac ig(n_i,[D_\mu n_j,D_\nu n_j]),
\label{de1}
\end{equation}
which is gauge invariant. By make using of the relation
\begin{equation}
(n_i,[n_j,A])=0,  \label{de3}
\end{equation}
one can prove that
\begin{equation}
f_{\mu \nu }^i=\partial _\mu A_\nu ^i-\partial _\nu A_\mu ^i+\frac ig%
(n_i,[\partial _\mu n_j,\partial _\nu n_j]),  \label{wu1}
\end{equation}
in which $A_\mu ^i$ is the projection of $A_\mu $ on the $n_i$
\[
A_\mu ^i=(A_\mu ,n_i).
\]

From the above discussion, one can find that this decomposition is
different from the work of Faddeev et al. With the idea that the
connection should possess its inner structure thus can be
decomposed, we express the connection $A_\mu$ with the covariant
part $b_\mu$ and the non-covariant part $a_\mu$, i.e.
$A_\mu=a_\mu+b_\mu$. In the case of SU(N), we have
\begin{equation}
a_\mu=(A_\mu,n_i)+\frac 1{ig}[\partial _\mu n_i,n_i], \;\;\;\;
b_\mu=-\frac 1{ig}[D_\mu n_i,n_i].
\end{equation}
In the section I, we discussed the non-covariant part $a_\mu$
should give the geometry information of system. In the following
section we can find that this part corresponds to the Wu-Yang
potential. In the paper of Faddeev et al., the decomposition of
SU(N) was given by introducing the new variables and they focused
on the covariant part $b_\mu$ to consider the effective
Lagrangian.

\section{Wu-Yang potential and the sourceless solution of the SU(N) gauge
field}

Since it was proposed, the existence of Wu-Yang potential has been
doubted for a long time. As a class of important gauge field
configurations for QCD, Wu-Yang monopoles are fundamental for
confinement and compete with the instanton-like configurations
which are responsible for chiral symmetry breaking \cite{Konishi}.
In this section we will be sure that the Wu-Yang potential exist
in the $SU(N)$ gauge theory.

From Eq. (\ref{loc1}) one have
\begin{equation}
\partial _\mu n_i=ig[\tilde{A}_\mu ,n_i],  \label{19}
\end{equation}
where $\tilde{A}_\mu =\frac 1{{ig}}\partial _\mu UU^{\dagger }$, is
obviously a flat connection, i.e.,
\begin{equation}
\partial _\mu \tilde{A}_\nu -\partial _\nu \tilde{A}_\mu -ig[\tilde{A}_\mu ,%
\tilde{A}_\nu ]=0  \label{flat1}
\end{equation}
In terms of Eq. (\ref{wu1}) and (\ref{flat1}) we can obtain
\[
\frac 1{ig}(n_i,[\partial _\mu n_j,\partial _\nu n_j])=\partial _\mu a_\nu
^i-\partial _\nu a_\mu ^i,
\]
where
\begin{equation}
a_\mu ^i=(\tilde{A}_\mu ,n_i)
\end{equation}
are the Wu-Yang potential. Thus we show that Wu-Yang potentials are Abelian
projection of the flat connection in the $SU(N)$ gauge theory. Furthermore,
since there are $(N-1)$ Abelian projection directions, one can find $(N-1)$
Wu-Yang potentials in $SU(N)$ gauge theory.

Finally, from Eq. (\ref{de1}) we note that when $D_\mu n_i=0,$%
\begin{equation}
f_{\mu \nu }^i=(F_{\mu \nu },n^i).  \label{30}
\end{equation}
But from the relation that
\begin{equation}
(D_\mu D_\nu -D_\nu D_\mu )n_i=[F_{\mu \nu },n_i],  \label{32}
\end{equation}
we find that $[F_{\mu \nu },n_i]=0\;(i=1,2,...,L).$ That is to say, when $%
D_\mu n_i=0,\;F_{\mu \nu }$ must commute with $n_i.$ Therefore using Eq. (%
\ref{30}), $F_{\mu \nu }$ can be expressed as
\begin{equation}
F_{\mu \nu }=f_{\mu \nu }^in_i.  \label{34}
\end{equation}
Then from $D_\mu n_i=0$, we have
\[
D_\nu F_{\mu \nu }=(\partial _\nu f_{\mu \nu }^i)n_i.
\]
Hence the solutions of equation
\begin{equation}
\partial _\nu f_{\mu \nu }^i=0,\;\;\;\;\;i=1,2,...,L,  \label{36}
\end{equation}
corresponding to the solutions of the sourceless equation
\[
D_\nu F_{\mu \nu }=0.
\]
If we adopt the Lorentz condition $\partial _\nu A_\nu ^i=0\;(i=1,2,...,L),$
from Eqs. (\ref{wu1}) and (\ref{36}) one can get
\begin{equation}
\nabla ^2A_\mu ^i=\partial _\nu K_{\mu \nu }^i,  \label{38}
\end{equation}
where
\[
K_{\mu \nu }^i=\frac ig(n_i,[\partial _\mu n_j,\partial _\nu n_j]).
\]
The solution of Eq. (\ref{38}) is
\begin{equation}
A_\mu ^i=-\frac 1{4\pi }\int G(x-x^{\prime })\partial _\nu K_{\mu \nu
}^i(x^{\prime })d^4x^{\prime },  \label{39}
\end{equation}
where $G(x-x^{\prime })$ is the retarded Green's function
\[
\nabla ^2G(x-x^{\prime })=-4\pi \delta ^4(x-x^{\prime }).
\]
Then, it follows from Eq. (\ref{con-de}) that the equation
\begin{equation}
A_\mu =-\frac 1{4\pi }[\int G(x-x^{\prime })\partial _\nu K_{\mu \nu
}^i(x^{\prime })d^4x^{\prime }]n_i+\frac 1{ig}[\partial _\mu n_i,n_i],
\label{40}
\end{equation}
is the sourceless solution of the gauge field equation.

\section{Conclusion}

In this paper, we give the decomposition of SU(2) and SU(N)
connection with the idea that connection should possess their
inner structures. One can find that the decomposition in this
paper have no any hypothesis or ansatz since the expressions of
connections are obtained with their definitions directly. In terms
of the decomposition of SU(2) connection, we give the 't Hooft
invariant gauge field tensor which can describe the magnetic
monopole correctly. And with the expression of SU(N) connection we
consider the Wu-Yang potential which is responsible for the
confinement in QCD.

\section{Acknowledgement}

This work was supported by the National Natural Science Foundation of China
and the Doctoral Foundation of the People's Republic of China.

\end{document}